\documentclass[useAMS,usenatbib]{mn2e}
\usepackage{epsf,palatino,url,graphicx,array,setspace,amsmath,amssymb,fancyhdr,multirow,lscape,appendix,rotating,gensymb}
\usepackage[usenames,dvipsnames,svgnames,table]{xcolor}
\def\apj{ApJ }
\def\aap{A\&A }

\def\mnras{MNRAS }

\def\apjs{ApJS }

\def\apjl{ApJL }

\def\apjs{ApJS }

\title[\emph{XMM} observations of Swift~J045106.8-694803]{\emph{XMM-Newton} observation of the highly magnetised accreting pulsar Swift~J045106.8-694803: Evidence of a hot thermal excess}
\author[E. S. Bartlett, M. J. Coe and  W. C. G. Ho]{E. S. Bartlett$^{1}$\thanks{E-mail:e.s.bartlett@soton.ac.uk (ESB)}, M. J. Coe$^{1}$,  W. C. G. Ho$^{2}$\\
$^{1}$School of Physics and Astronomy, University of Southampton, Highfield, Southampton, SO17 1BJ, United Kingdom\\
$^{2}$School of Mathematics, University of Southampton, Southampton, SO17 1BJ, United Kingdom}

\begin{document}

\date{Accepted 2013 September 10.  Received 2013 September 9; in original form 2013 February 4}

\pagerange{\pageref{firstpage}--\pageref{lastpage}} \pubyear{2013}

\maketitle

\label{firstpage}

\begin{abstract}

Several persistent, low luminosity ($L_X\sim10^{34}$~erg~s$^{-1}$), long spin period (P\textgreater100~s) High Mass X-ray Binaries have been reported with blackbody components with temperatures \textgreater1~keV. These hot thermal excesses have correspondingly small emitting regions (\textless1~km$^{2}$) and are attributed to the neutron star polar caps. We present a recent \emph{XMM-Newton} target of opportunity observation of the newest member of this class, Swift~J045106.8-694803. The period was determined to be 168.5$\pm$0.2~s as of 17 July 2012 (MJD = 56125.0). At $L_X\sim10^{36}$~erg~s$^{-1}$, Swift~J045106.8-694803 is the brightest member of this new class, as well as the one with the shortest period. The spectral analysis reveals for the first time the presence of a blackbody with temperature $kT_{BB}=1.8^{+0.2}_{-0.3}$~keV and radius $R_{BB}=0.5\pm0.2$~km. The pulsed fraction decreases with increasing energy and the ratio between the hard (\textgreater2~keV) and soft (\textless2~keV) light curves is anticorrelated with the pulse profile. Simulations of the spectrum suggest that this is caused by the pulsations of the blackbody being  $\sim\pi$ out of phase with those of the power law component. Using a simple model for emission from hot spots on the neutron star surface, we fit the pulse profile of the blackbody component to obtain an indication of the geometry of the system.

\end{abstract}

\begin{keywords}
X-rays: binaries, stars: emission line, Be, neutron, pulsars: individual: Swift~J045106.8-694803
\end{keywords}

\section{Introduction}

Be/X-ray binaries (BeXRBs) are stellar systems in which a compact object, almost exclusively a neutron star, orbits a main sequence Be star. These stars are rotating rapidly, causing an enhancement in the equatorial material which in turn, leads to hydrogen emission lines in the spectrum. This is a transient phenomenon and thus any star that has exhibited hydrogen emission lines in its spectrum at some time is classed as a Be star. The compact object orbits the primary in a highly eccentric orbit and accretes matter from the equatorial outflow. There are two types of outburst associated with the X-ray emission of BeXRBs: Type I outbursts have L$_X$ in the range 10$^{36}$-10$^{37}$ erg~s$^{-1}$ and occur periodically around the time of the periastron passage of the neutron star. Type II outbursts reach higher luminosities, L$_X \geq$10$^{37}$ erg~s$^{-1}$, last much longer and show no correlation with orbital phase \citep{Stella86}. These are thought to be caused by an enhancement of the circumstellar disc allowing accretion to occur at any phase of the orbit at a much higher rate. For a review of the observational properties of BeXRBs see \citet{Reig11}.

\begin{table*}
 \centering
\caption{Summary of sources with $kT_{BB}$\textgreater1.0~keV. $R_{BB}$ is the radius of the emitting region implied by $L_X$ and $kT_{BB}$. $D$ is the assumed distance to the source in kpc, errors are shown when they are accounted for in the calculation of the blackbody radius. All errors are 90\% confidence level.}\label{tab:BeX}
{\scriptsize\begin{tabular}{lccccccl}
\hline\hline\noalign{\smallskip}
\multirow{2}{*}{Source} & Period & $kT_{BB}$ & $L_X$ & $R_{BB}$ & $D$ & Energy Range &\multirow{2}{*}{Reference} \\\noalign{\smallskip}
 & s & keV & erg~s$^{-1}$ & m & kpc &  keV & \\\noalign{\smallskip}
\hline\noalign{\smallskip}
RX~J0146.9+6121 & $1396.1\pm0.3$ & $1.11^{+0.07}_{-0.06}$ & $\sim1.5\times10^{34}$ & $140^{+20}_{-10}$ & 2.5 & 0.3--10.0 & \citet{LaPalombara06} \\
\multirow{2}{*}{X Persei} & $839.3\pm0.3$ & $1.35\pm0.03$ & $\sim1.4\times10^{35}$ & $361\pm3$ & 1 &  0.3--10.0 & \citet{LaPalombara07} \\
 & $\sim837$ & $1.45\pm0.02$ & $\sim4.2\times10^{34}$ & $130$ & 0.95 & 3.0-10.0 & \citet{Coburn01} \\
RX~J1037.5-5647 & $853.4\pm0.2$ & $1.26_{-0.09}^{+0.16}$ & $\sim1.2\times10^{34}$ & $130^{+10}_{-20}$ & 5 & 0.2--10.0 & \citet{LaPalombara09} \\
RX~J0440.9+4431 & $204.96\pm0.02$ & $1.34\pm0.04$ & $\sim8\times10^{34}$ & $270\pm20$ & 3.3 & 0.3--12.0 & \citet{LaPalombara12} \\
SXP~1062 & $\sim1062$ & $1.54\pm0.16$ & $2.5\pm0.5\times10^{35}$ & $190_{-40}^{+60}$ & 55 & 0.2--12.0 & \citet{Henault11} \\
4U~2206+54 & $5593\pm10$ & $1.63\pm0.03$ & $\sim3\times10^{35}$ & $370\pm40$ & 2.6 & 0.2--12.0 & \citet{Reig12} \\
Swift~J045106.8-694803 & $168.5\pm0.2$ & $1.8_{-0.3}^{+0.2}$ & $(9.8\pm0.9)\times10^{35}$ & $500\pm200$ & $50.6\pm2.1$ & 0.2--10.0 & this work \\
\hline
\end{tabular}}\newline
\end{table*}
BeXRBs are the most numerous subclass of HMXB and have been predominately detected via the pulsations of the neutron star (e.g. \citealt{Galache08}). The number of known HMXBs has increased dramatically since the launch of satellites such as the \emph{R\"{o}ntgen Satellite} (\emph{ROSAT}, \citealt{Truemper82}) and the \emph{Rossi X-ray Timing Explorer} (\emph{RXTE}, \citealt{Bradt93}) particularly in the Magellanic Clouds \citep{Liu00,Liu05}. Given the large sample size of objects now available, we are able to study the properties of these objects on a statistically significant scale. The X-ray spectra of BeXRBs are characterised by intrinsically absorbed power laws with a photon indices, $\Gamma$, in the range 0.6-1.4 \citep{Haberl08}, with high energy cut-offs in the range 10-30~keV \citep{Lutovinov05,Reig11}.

Some authors have reported a soft excess in the spectra of HMXB pulsars, with blackbody temperatures $kT_{BB}$\textless0.5~keV (for e.g. see \citealt{Hickox04}). \citet{Hickox04} suggest that a soft excess is in fact present in most, if not all, HMXB spectra, though not always detected due to the high intrinsic absorption and flux of some sources. For systems with $L_X\gtrsim10^{38}$~erg~s$^{-1}$, this excess is thought to originate from the reprocessing of hard X-rays, most likely at the inner radius of the accretion disc surrounding the neutron star. For less luminous sources ($L_X\lesssim10^{36}$~erg~s$^{-1}$), the soft excess is attributed to other processes, e.g. thermal emission from the neutron star's surface. HMXBs of intermediate luminosity can show emission from either or both types of soft excess \citep{Hickox04}.

Recent observations with \emph{XMM-Newton} has revealed that a handful of BeXRBs have blackbody components with $kT_{BB}$ in excess of 1~keV and a derived emitting region $R<R_{NS}$ (see Table \ref{tab:BeX}). Such a small radius indicates emission from a hot spot on the neutron star, possibly from the magnetic polar cap. These sources all have low level X-ray emission ($L_X\sim10^{34-35}$~erg~s$^{-1}$) and long pulse periods (P\textgreater100~s).

Here we report on \emph{XMM-Newton} observation of another possible member of this group of BeXRBs: Swift~J045106.8-694803. This source was detected in the Large Magellanic Cloud (LMC) by the \emph{Swift}/BAT hard X-ray survey \citep{atel} and was followed by a 15.5~ks observation with the \emph{Swift} XRT instrument. This confirmed the position of the source and revealed a periodic signal at 187~s. From the accretion model of \citet{Ghosh79}, \citet{Klus12} derived a magnetic field $B\sim1.2\times10^{14}$~G from the spin-up rate, indicating that Swift~J045106.8-694803 is a highly magnetised accreting pulsar (i.e., neutron star with a super strong magnetic field $B\gtrsim10^{14}$~G). However, there are several interpretations of the high spin down rates observed in these sources which do not require super strong magnetic fields, such as accretion of magnetised material (e.g. \citealt{Ikhsanov12, Ikhsanov13}) or quasi-spherical subsonic accretion (e.g. \citealt{Shakura12}).

\section{Observations and Data Reduction}

\begin{table}
\centering
\caption{\emph{XMM-Newton} EPIC observations of Swift~J045106.8-694803 on 2012 July 17}\label{table:obs}
{\scriptsize\begin{tabular}{ccccccc}
\hline\hline\noalign{\smallskip}
Camera & Filter & Read out & \multicolumn{3}{c}{Observation} & Exp. \\\noalign{\smallskip}
& & Mode & Date (MJD) & Start(UT) & End(UT) & (ks) \\\noalign{\smallskip}
\hline\noalign{\smallskip}
MOS1/2 & Medium & Full Frame & \multirow{2}{*}{56125.0} & 00:39 & 03:03 & 8.6 \\\noalign{\smallskip}
pn & Medium & Full Frame & & 01:01 & 02:59 & 7.0 \\\noalign{\smallskip}
\hline
\end{tabular}}
\end{table}A $\sim$7~ks \emph{XMM-Newton} target of opportunity (ToO) observation was performed during satellite revolution \#2308, MJD = 56125.0 (2012 July 17). Data from the European Photon Imaging Cameras were processed using the \emph{XMM-Newton} Science Analysis System v11.0 (SAS) along with software packages from FTOOLS v6.12. Table \ref{table:obs} 
summarises the details of the EPIC observations.

The MOS \citep{Turner01} and pn \citep{Struder01} observational data files were processed with \textsf{emproc} and \textsf{epproc} respectively. The data were screened for periods of high background activity by examining the \textgreater10~keV count rate. The pn and MOS count rates were below the recommended filtering threshold for the duration of the observation and so no filter was applied. The final cleaned pn image included ``single'' and ``double'' (PATTERN$\leq$4) pixel event patterns in the 0.2--10.0~keV energy range. ``Single'' to ``quadruple ''(PATTERN$\leq$12) pixel events were selected for the cleaned MOS images in the same energy range. Photon arrival times were converted to barycentric dynamical time, centred at the solar system barycenter, using the SAS task \textsf{barycen}.

Images, background maps and exposure maps were created for all detectors in the 0.2--10.0~keV energy range. A box sliding detection was performed simultaneously on all 3 images twice (the first with a locally estimated background the second using the background map) with the task \textsf{eboxdetect}, followed by the maximum likelihood fitting using the task \textsf{emldetect}. This process resulted in a list of sources including their positions, errors and background subtracted counts. 

Source counts were extracted from a circular region with radius 61\arcsec, as recommended by the SAS task \textsf{eregionanalyse} which calculates the optimal radius for the source extraction by maximising the signal to noise. Background counts were extracted from a region of identical size offset from the source. This region falls on a neighbouring CCD in the pn detector and on the same chip in the MOS1 and MOS2 detectors. The background subtraction was performed using the \textsf{epiclccorr} task which also corrects for bad pixels, vignetting and quantum efficiency.

Source and background spectra were extracted from the same regions. Again, ``single'' and ``double'' pixel events (PATTERN$\leq$4) were accepted for the pn detector with all bad pixels and columns disregarded (FLAG=0). For the MOS spectra, ``single'' to ``quadruple ''(PATTERN$\leq$12) pixel events were selected with quality flag \#XMMEA\_EM. The area of source and background regions were calculated using the \textsf{backscal} task. Response matrix files were created for each source using the task \textsf{rmfgen} and \textsf{arfgen}.

\section{Analysis and Results}

\subsection{Position}

\begin{figure}
 \centering
  \includegraphics[width=0.5\textwidth]{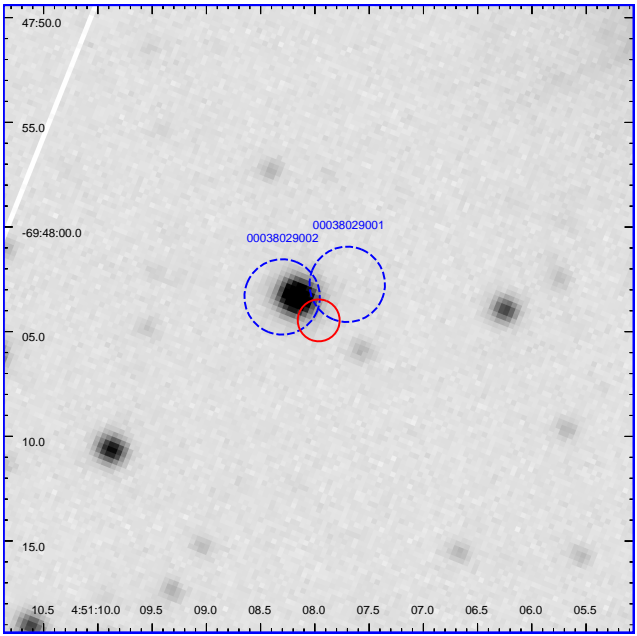}
   \caption{$V$-band image of Swift~J045106.8-694803, taken with EFOSC2 on the NTT at La Silla, Chile with the \emph{XMM-Newton} (solid red) and \emph{Swift} (broken blue) $1\sigma$ error circles. The two \emph{Swift} observation IDs are labeled.}\label{fig:posn}
\end{figure}

The simultaneous source detection performed on the three EPIC cameras determined the position of Swift~J045106.8-694803 as RA(J2000)=$04^h51^m06.7^s$ Dec(J2000)=$-69\degree48\arcmin04.2\arcsec$. The $1\sigma$ systematic uncertainty was assumed to be 1\arcsec in accordance with the findings of the \emph{XMM-Newton} Serendipitous Source catalogue \citep{2XMM}. This is an order of magnitude larger than the statistical error derived in the source detection, and as such is the dominant error on the position. This position is consistent with the \emph{Swift} positions reported by \citet{atel} and \citet{Klus12} confirming that these three detections are the same source. Figure \ref{fig:posn} shows a \emph{V}-band image with the location of the \emph{Swift} and \emph{XMM} positions with radii equal to the $1\sigma$ errors. The images was taken with the ESO Faint Object Spectrograph and Camera (EFOSC2) mounted at the Nasmyth B focus of the 3.6m New Technology Telescope (NTT), La Silla, Chile on the night of 2011 December 9 (MJD = 55904). The optical counterpart of Swift~J045106.8-694803 is [M2002] 9775 \citep{Massey02}, located at RA(J2000)=$04^h51^m06.96^s$ $-69\degree48\arcmin03.0\arcsec$.

\subsection{Timing Analysis}

\begin{figure*}
 \centering
  \includegraphics[angle=90,width=1.0\textwidth]{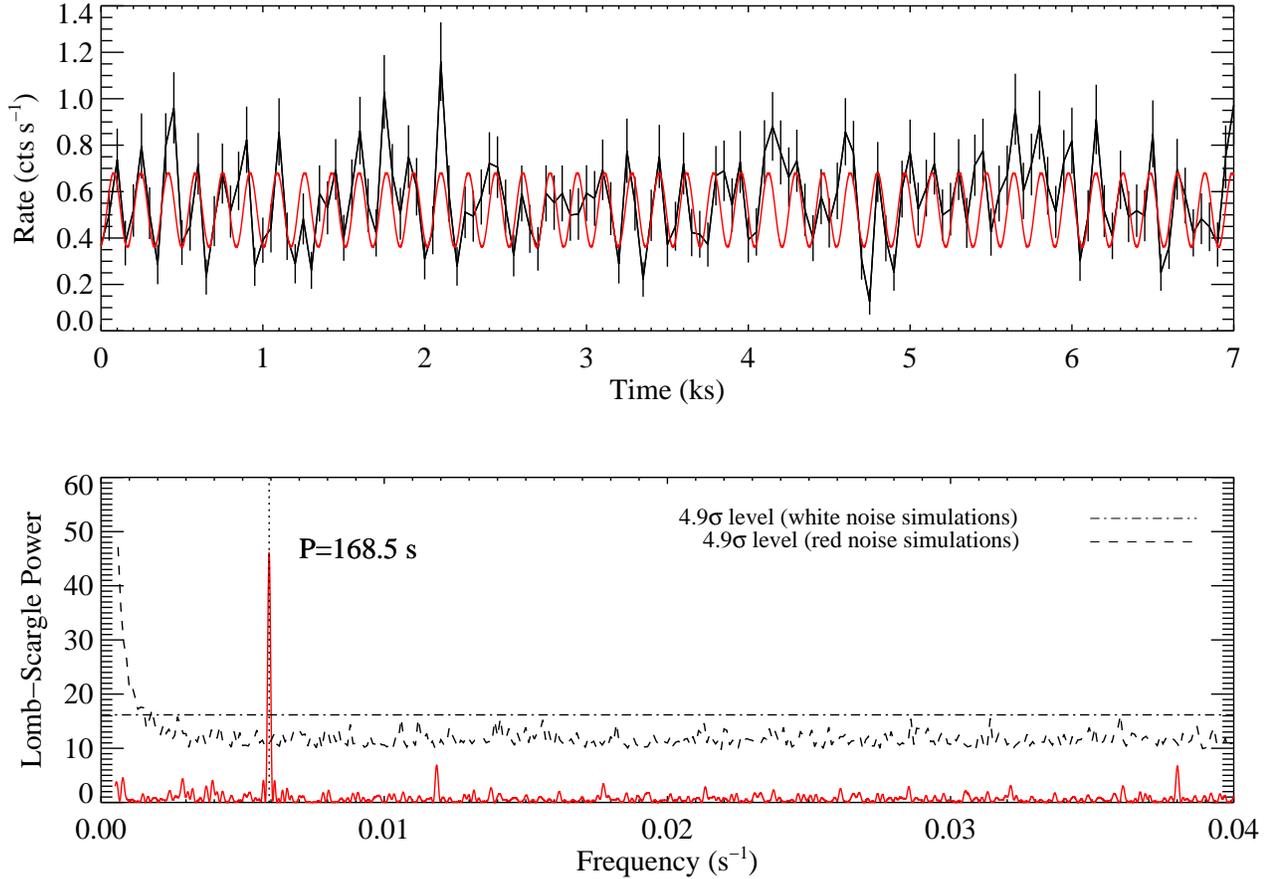}
  \caption{Top panel shows the EPIC pn light curve of Swift J045106.8-694803 with 50~s bins. Bottom panel shows the Lomb-Scargle periodogram with the detected period marked along with the 99.9999\% significance levels determined by white and red noise simulations.}\label{Fig:LombScargle}
\end{figure*}

Figure \ref{Fig:LombScargle} shows the light curve of Swift J045106.8-694803. The bottom panel shows the Lomb-Scargle periodogram of the light curve with a bin time of 20~s. A sine wave with the detected period was fit to the 20~s bin light curve and is overlaid for clarity. A period at 168.8~s with a power of 46.1 was detected rising to 82.7 when the bin size of the light curve is reduced to 0.1~s. Monte Carlo simulations with both red and white noise light curves were performed to determine the significance of this detection. A bin time of 20~s was also employed for the simulations to reduce the processing time. One million white noise light curves were generated by ``scrambling'' the original light curve (i.e. reassigning the flux values to different time stamps) using a random number generator. This method makes no assumption about the underlying distribution of the light curve. Lomb-Scargle analysis was performed on each of these light curves and the highest power recorded. None of the 1,000,000 light curves generated produced a peak in the periodogram greater than 17.0. This suggests that the period discovered in the light curve has a significance \textgreater 99.9999\%  or $4.9\sigma$.

One million light curves were generated with a power law slope of -2.0 and the same statistical properties (mean, standard deviation and bin time) as the EPIC-pn light curve, using the method of \citet{Timmer95}\footnote{using the Interactive Data Language (IDL) procedure \textsf{rndpwrlc.pro}}. The light curves were initially simulated with a duration ten times longer than that of the actual data and were then cut down to the observed duration to minimise the effect of red noise leakage. Gaussian noise was added to each point of this new light curve by drawing a random deviate from a Gaussian distribution with mean and variance equal to each data point following the method detailed by \citealt{Uttley03}). Any bins with a negative count rate were set to zero. Lomb-Scargle analysis was performed on each of the simulated time series. Unlike the white noise simulations, the significance of a peak depends on the frequency. The broken line in Fig \ref{Fig:LombScargle} shows the 99.9999\% significance contour. Both the white and red noise simulations indicate that this period is significant.

The error in the period was estimated by varying the original light curve within the errors on each data point, using a Gaussian random number generator, 10,000 times. As with the simulations to determine the significance of the detection, Lomb-Scargle analysis was performed on each of these light curves. To speed up the processing time of the simulation, the light curve was only searched for periods between 50~s and 2000~s. The resulting histogram is well fit by a Gaussian with mean 168.5~s and a standard deviation of 0.16~s. As such we determine the period of Swift J045106.8-694803 to be 168.5$\pm$0.2 as of MJD = 56125.0.\begin{figure*}
 \centering
  \includegraphics[width=0.9\textwidth]{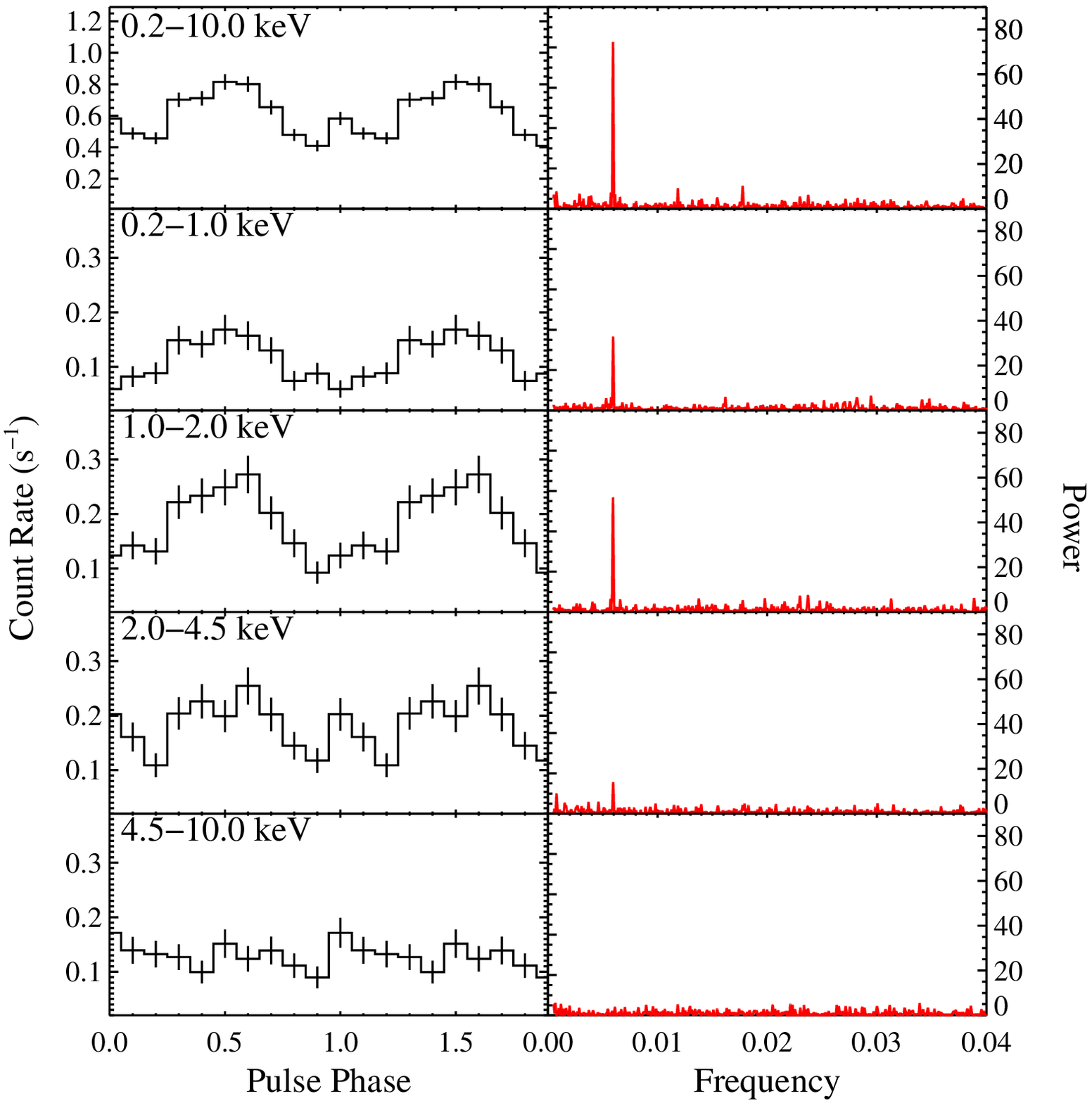}
  \caption{Left panels show the background subtracted pulse profiles from the EPIC-pn detector, folded on the 168.5~s period detected. Right panels show the Lomb-Scargle periodogram in the same energy range.}\label{fig:pulse}
\end{figure*}

The light curve was split into four energy ranges, 0.2-1.0~keV, 1.0-2.0~keV, 2.0-4.5~keV and 4.5-10.0~keV with approximately equal count rates (0.12, 0.18, 0.17 and 0.13 counts~ s$^{-1}$ respectively). Figure \ref{fig:pulse} shows the pulse profiles for each of these light curves and the entire 0.2-10.0~keV light curve, each normalised to the average count rate in the energy range. The zero phase point was determined from the phase shift found from the sine wave fit to the 0.2-10.0~keV light curve (see above). Lomb-Scargle analysis was performed on each of the light curves.

The strongest detection of the period is in the 1.0-2.0~keV energy range, with a Lomb-Scargle power of 56.7. This does not appear to be an issue relating to photon counting statistics, as the period is barely detected in the 2.0-4.5~keV energy range (Lomb-Scargle power of 15.2) which has an almost identical average count rate. The strongest period identified in the 4.5-10.0~keV was at 11.3~s with a Lomb-Scargle power of 7.8, this is likely to be noise rather than the detection of a second period.

We investigated whether the lack of significant pulsations at higher energies could be due to a change in the shape of the pulse profile, as Lomb-Scargle analysis is more sensitive to sinusoidal variations. We checked for periods using the epoch folding methods of \citet{Leahy87}. The lightcurve is folded on each trial period and tested to see if it is consistent with a constant count rate with a $\chi^2$ test. This reinforces the results from the Lomb-Scargle analysis, with the strongest detection in the 1.0-2.0~keV range and no detection in the 4.5-10.0~keV range. The first harmonic of the period was the strongest period identified in the 2.0-4.5~keV light curve, suggesting the pulsed profile may become double peaked at higher energies.

The pulsed fraction of each light curve was calculated by fitting a sine wave with the period fixed at the value found in the full 0.2-10.0~keV energy range. The phase, amplitude and the average value of the light curve were allowed to vary and the ratio of the amplitude and average value were taken. This is equivalent to taking the ratio of the difference of the maximum and minimum value of the sine wave with the sum of these values. This parameter can vary between 1 (completely pulsed) and 0 (constant rate). The values range from 0.47$\pm$0.05 for the 1.0-2.0~keV energy range down to 0.08$\pm$0.06 for the 4.5-10.0~keV range. The epoch folding suggests that the profile becomes double peaked at higher energies. The fit was also performed with the period fixed at the second harmonic for the last two energy bands. The results are summarised in Table \ref{tab:frac}.
\begin{table}
 \centering
\caption{Summary of timing results. Pulsed fraction for the 4.5-10.0~keV energy range are the $3\sigma$ upper limit}\label{tab:frac}
\begin{tabular}{lccc}
 \hline\hline
Energy & Lomb-Scargle & \multicolumn{2}{c}{Pulsed Fraction} \\\noalign{\smallskip}
Range (keV) & Power & P=168.5~s & P=84.3~s \\\noalign{\smallskip}
\hline\noalign{\smallskip}
0.2-10.0 & 82.7 & 0.43$\pm$0.03 & - \\\noalign{\smallskip}
0.2-1.0 & 36.1 & 0.42$\pm$0.07 & - \\\noalign{\smallskip}
1.0-2.0 & 56.7 & 0.47$\pm$0.05 &  - \\\noalign{\smallskip}
2.0-4.5 & 15.2 & 0.34$\pm$0.05 & 0.13$\pm$0.06\\\noalign{\smallskip}
4.5-10.0 & - & $<0.26$ & $<0.32$\\
\hline
\end{tabular}
\end{table}

We consider the light curves for two energy ranges, 0.2-2.0~keV and 2.0-10.0~keV, with equal count rates ($0.293\pm0.007$ and $0.305\pm0.007$ respectively). The hardness ratio ($HR$) between the ``soft'' (\textless2~keV) and ``hard'' (\textgreater2~keV) light curves was calculated using the formula:\begin{equation}
 HR=\frac{C_{hard}-C_{soft}}{C_{hard}+C_{soft}}
\end{equation} where $C_{hard}$ and $C_{soft}$ are the count rates in the hard and soft bands respectively. The hardness ratio can vary between -1.0 (zero counts in the 2.0-10.0~keV band) and 1.0 (zero counts in the 0.2-2.0~keV band).\begin{figure}
 \centering
  \includegraphics[angle=90,width=0.47\textwidth]{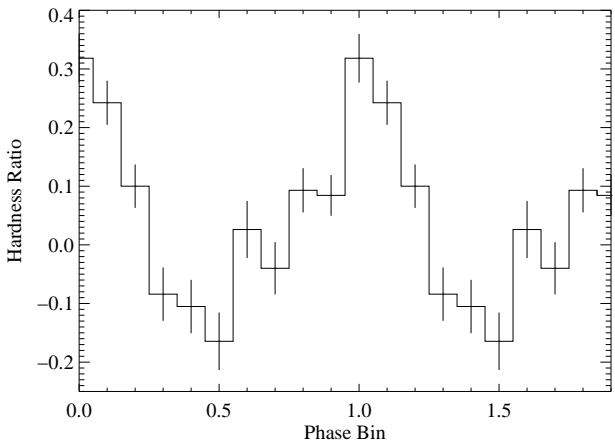}
  \caption{Hardness ratio, $\frac{C_{hard}-C_{soft}}{C_{hard}+C_{soft}}$, as a function of pulse phase. $C_{soft}$ is the 0.2-2.0~keV count rate and $C_{hard}$ is the 2.0-10.0~keV count rate. The phase shown is the same as that of the pulse profiles shown in Fig \ref{fig:pulse}.}\label{fig:HR}
\end{figure} Figure \ref{fig:HR} shows how the hardness ratio varies with pulse phase. From a comparison with the left panels of Fig. \ref{fig:pulse}, a clear anti-correlation between the hardness ratio and the pulse profile is evident, with the source getting harder with decreasing luminosity.

\subsection{Spectral Analysis}

\begin{table*}
\centering
\caption{Best fit parameters for the spectral fits. In all three models the \emph{phabs} component is fixed at $8.4\times10^{20}$~cm$^{-2}$ \citep{DL91}. Errors, where reported, are the 90\% confidence level. The codes for the different model components are: ``\emph{ph}'' for \emph{phabs}, ``\emph{vph}'' for \emph{vphabs}, ``\emph{po}'' for \emph{powerlaw}, ``\emph{bb}'' for \emph{blackbody}, ``\emph{di}'' for \emph{diskbb} and \emph{bmc} for bulk motion comptonisation.}\label{Tab:para}
\begin{tabular}{lcccccccc}
\hline\hline\noalign{\smallskip}
\multirow{2}{*}{Model} & $N_{H,i}$ & $\Gamma$ & normalisation$^{(a)}$ & $kT_{BB}$ & normalisation$^{(b)}$ & Flux$^{(c)}$ & $L_X^{(d)}$ & \multirow{2}{*}{$\chi_r^2$/dof}\\
&  [$10^{21}~cm^{-2}$]  & &  & [keV] &  &  &  & \\
\hline\noalign{\smallskip}
\emph{ph*vph*po} & $1.3\pm0.4$ & 0.97$\pm$0.05 & $2.0\pm0.1\times10^{-4}$ & - & - & $3.4_{-0.2}^{+0.1}$ & $1.03\pm0.09$ & 1.15/236\\\noalign{\smallskip}
\emph{ph*vph*(bb+po)} & $1.6_{-0.8}^{+1.0}$ &  $1.4_{-0.3}^{+0.5}$ & $1.9_{-0.2}^{+0.3}\times10^{-4}$ & $1.8_{-0.3}^{+0.2}$ & $1.9_{-1.1}^{+0.9}\times10^{-5}$ & $3.2_{-0.2}^{+0.1}$ & $0.98\pm0.09$ & 1.08/234\\\noalign{\smallskip}
\emph{ph*vph*(di+po)} & $5_{-5}^{+3}$ & $4_{-4}^{+1}$ & $1\pm1\times10^{-4}$ & 4.2$_{-0.6}^{+1.0}$ & $7\pm4\times10^{-4}$ & $3.0\pm0.1$ & $0.92\pm0.06$ & 1.09/234\\\noalign{\smallskip}
\hline\noalign{\smallskip}
\multirow{2}{*}{Model} & $N_{H,i}$ & $kT$ & $\alpha$ & $f$ & normalisation & $kT_{BB}$ & normalisation$^{(a)}$& \multirow{2}{*}{$\chi_r^2$/dof}\\\noalign{\smallskip}
& [$10^{21}~cm^{-2}$]  & [keV] & & & & [keV] & & \\\noalign{\smallskip}
\hline\noalign{\smallskip}
\emph{ph*vph*bmc} & $4_{-2}^{+3}$ & $0.11_{-0.02}^{+0.01}$ & $0.04_{-0.04}^{+0.07}$ & $ 1_{-2}^{+1}$ & $4_{-2}^{+10}\times10^{-5}$ & - & - & 1.14/234\\\noalign{\smallskip}
\emph{ph*vph*(bb+bmc)} & $2\pm1$ & $0.020_{-0.004}^{+0.025}$ & $0.6\pm0.5$ & $<-0.3$ & $0.2_{-0.1}^{+1.6}$ & $1.8\pm0.2$ &  $2.4_{-1.3}^{+0.7}\times10^{-5}$ & 1.08/232\\\noalign{\smallskip}
\hline
 \end{tabular}
\newline
\flushleft{\textbf{Notes.} $^{(a)}$ Defined in \textsf{XSPEC} as photons~keV$^{-1}$~cm$^{-2}$~s$^{-1}$ at 1~keV. $^{(b)}$ Units are 10$^{37}$~erg~s$^{-1}$~kpc$^{-2}$. Defined in \textsf{XSPEC} as $L_{39}/D_{10}$, where $L_{39}$ is the source luminosity in units of $10^{39}$~erg~s$^{-1}$, $D_{10}$ is the distance to the source in units of 10~kpc. $^{(c)}$ Observed flux in the 0.2-10.0~keV range. Units are 10$^{-12}$~erg~cm$^{-2}$~s$^{-1}$. $^{(d)}$ Source intrinsic luminosity in the 0.2-10.0~keV range, corrected for absorption and assuming a distance to the LMC of 50.6~kpc. Units are 10$^{36}$~erg~s$^{-1}$.}
\end{table*} The spectral analysis discussed here was performed using \textsf{XSPEC} \citep{Arnaud96} version 12.7.0. The three spectra from the different EPIC detectors were fit simultaneously with the models reported here plus an additional constant factor to account for the variations in the different detectors. The model parameters were constrained to be identical across the three instruments. The photoelectric absorption was split into two components. One, $N_{H,Gal}$, to account for the Galactic foreground extinction, fixed to $8.4\times10^{20}$~cm$^{-2}$ \citep{DL91} with abundances from \citet{Wilms00}, and a separate column density, $N_{H,i}$, intrinsic to the LMC with abundances set to 0.5 for elements heavier than helium \citep{Russel92} and allowed to vary. All errors are evaluated at the 90\% confidence level.\begin{figure}
\centering
\hspace{-50pt}
\includegraphics[angle=-90,width=0.56\textwidth]{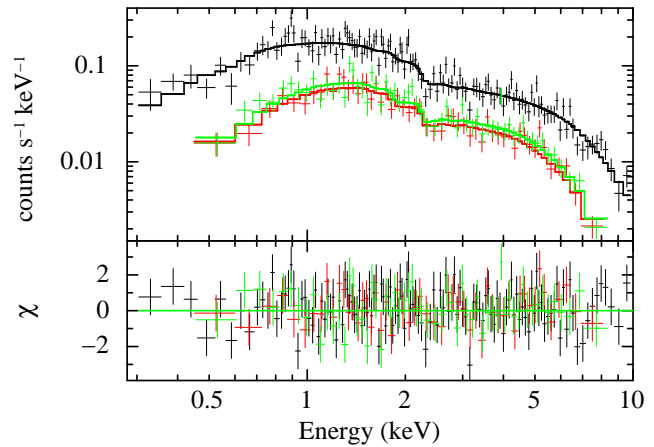}
\caption{The 0.2--10.0~keV EPIC-pn (black), EPIC-MOS1 (red), EPIC-MOS2 (green) spectra of Swift J045106.8-694803. Top panel displays the background subtracted spectrum with best fit \emph{phabs*vphabs(powerlaw+bbody)} model, bottom panel shows the residuals.}\label{fig:spec}\end{figure}

The spectra obtained from the three instruments were initially fit with a simple absorbed power law model (\emph{phabs*vphabs*powerlaw} in \textsf{XSPEC}). This led to an acceptable fit, with a $\chi^2$ of 270.5 for 236 degrees of freedom (dof) with a photon index of $\Gamma=0.97\pm0.05$ and intrinsic absorption $N_{H,i}=(1.3\pm0.4)\times10^{21}$~cm$^{-2}$. The photon index is typical for those seen in other BeXRBs, particularly in the SMC \citep{Haberl08}. 

The possibility of a thermal component was also explored and modeled with both a blackbody and diskbb model. Including these parameters improved the fit marginally ($\chi^{2}$ of 252.8 and 255.7 for 234 dof respectively) but F-tests suggest that these are significant at 99.96\% and 99.86\% respectively (i.e. \textgreater$3\sigma$ ). Fig. \ref{fig:spec} shows the 0.2--10.0~keV spectrum along with the best fit model (\emph{phabs*vphabs(powerlaw+bbody)}). The parameters of all the models discussed here are included in Table \ref{Tab:para}. The best fit parameters for the \emph{phabs*vphabs*(diskbb+powerlaw)} are both unphysical and poorly constrained (e.g. $\Gamma=4_{-4}^{+1}$ is extremely soft and $kT=4.2_{-0.6}^{+1.0}$~keV is too hot for a disc, which have $kT\sim0.1$~keV), as such, only the results of the \emph{phabs*vphabs*(bbody+powerlaw)} model are discussed in detail. The total unabsorbed flux from the blackbody component is $1.3\pm0.8\times10^{-12}$~erg~cm$^{-2}$~s$^{-1}$, accounting for approximately 40\% of the total emission of the source.

An intrinsically narrow Gaussian was added to the model at 6.4~keV to see if any evidence for an Fe-K$\alpha$ line exists. The upper limit on the equivalent width of this component was derived as 0.3~keV. Allowing the energy or width of this feature to vary does not alter this result.

The spectrum was fit with the self consistent Bulk Motion Comptonisation model (e.g. \citealt {Borozdin99}; \emph{phabs*vhabs*bmc} in \textsf{XSPEC}) in the same manner as above. This model is characterised by a thermal temperature, $kT$, which represents the temperature of the soft seed photons; an illumination parameter, $f$; and a spectral index $\alpha$. The best fit parameters are shown in Table \ref{Tab:para}. This model has a $\chi^2$ value of 266.9 for 234 dof, i.e. a $\chi^2_r$ value similar to that of the original absorbed power law model. The spectral energy index is abnormally low ($0.04_{-0.04}^{+0.07}$), where normal is anything between 0.6 and 1.4 \citep{Haberl08}. When a blackbody was added to the model (\emph{phabs*vphabs(bbody+bmc)} in \textsf{XSPEC} the $\chi^2$ fell to 249.8 for 232 dof. Whilst several parameters are still poorly constrained (e.g. $f<-0.3$), the spectral energy index is consistent with those reported for BeXRBs ($\alpha=0.6\pm0.5$). The blackbody temperature is identical to that reported for the \emph{bbody+powerlaw} model ($kT_{BB}=1.8\pm0.2$).

Figure\begin{figure}
\centering
\includegraphics[angle=90,width=0.5\textwidth]{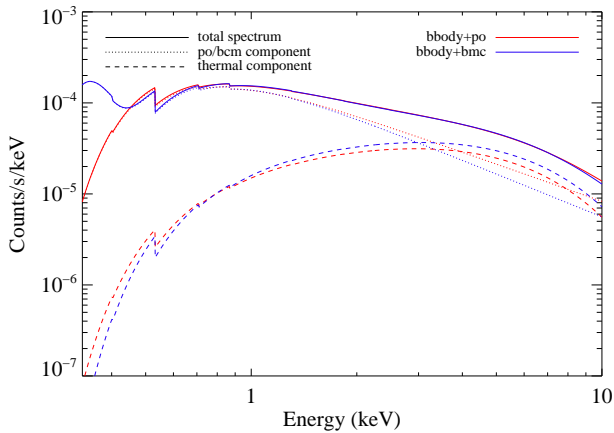}
\caption{Figure shows a comparison between the \emph{bbody+powerlaw} and the \emph{bbody+bmc} model over the energy range 0.3-10.0~keV.}\label{fig:compare}
\end{figure} \ref{fig:compare} Shows a comparison between the \emph{bbody+powerlaw} model and the \emph{bbody+bmc} model. Between 0.4-10.0~keV the two models are indistinguishable. Importantly, the phenomenological \emph{powerlaw} component approximates the physical \emph{bmc} component at energies $\gtrsim0.5$~keV. Since the \emph{bbody+bmc} is poorly constrained and to allow for easier comparison with previous works, we will use values determined by the simple, empirical \emph{bbody+powerlaw} model.

\subsection{Modelling the Phase Resolved Spectra and Pulse Profiles}

The anticorrelation seen between the hardness ratio and pulse profile has previously been reported for another persistent BeXRB, RX~J0146.9+6121 by \citep{LaPalombara06}. Pulsed phased spectroscopy revealed that the change in spectra could be explained with a static blackbody and variable power law (among other solutions). This possibility was explored by generating 10,000 EPIC-pn spectra with the same absorption, photon index and blackbody temperature as the best fit model using the \textsf{XSPEC} command \textsc{fakeit}. These parameters were fixed as they are linked to the physical properties of the system and/or the local environment and so are unlikely to change on a timescale of seconds. The value of the normalisation for the power law and blackbody varied from zero to $5.0\times10^{-4}$ in steps of $5.0\times10^{-6}$. The total number of counts as well as the hardness ratio was calculated for each of the simulated spectra. The results of the simulation were searched for the normalisation values which could reproduce the top two panels in Fig. \begin{figure}
\centering
\includegraphics[width=0.5\textwidth]{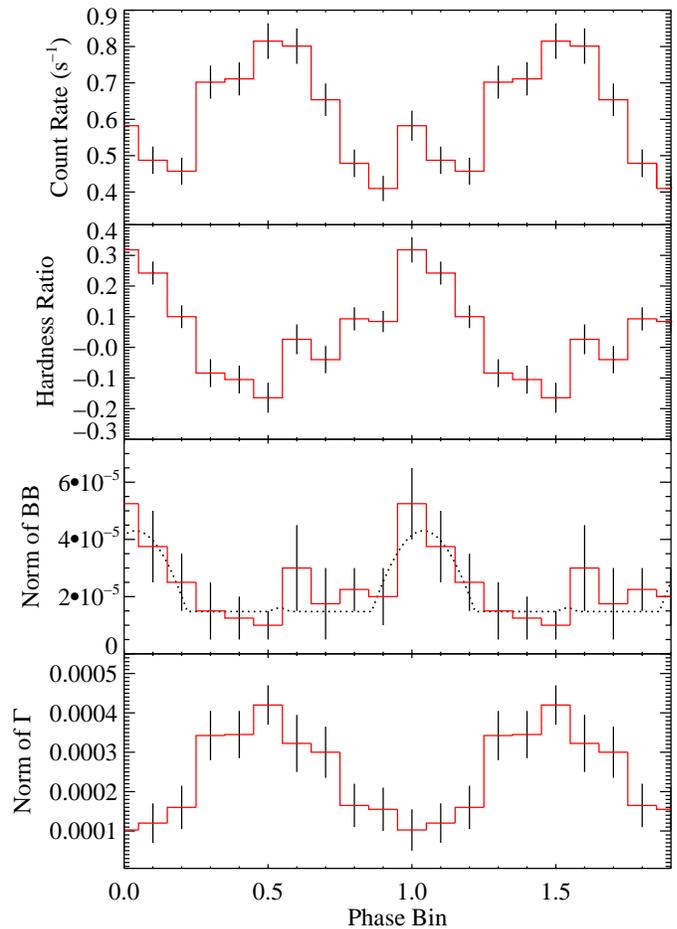}
\caption{Panels from top to bottom show (1) 0.2-10.0~keV pulse profile (2) Hardness ratio (3) The normalisation of the blackbody (BB) required to produce the hardness ratio and count rate of the given phase bin and (4) The normalisation of the power law ($\Gamma$) required for the given phase bin. The broken line in panel (3) is the best-fit pulse profile (see text)}\label{fig:sims}
\end{figure}\ref{fig:sims} within errors, i.e. the same count rate and hardness ratio as each of the phase bins.

The results of the simulations, along with the pulse profile and hardness ratio, are shown in Fig. \ref{fig:sims}. Interestingly, a constant blackbody could not reproduce the range of hardness ratios seen. A modulation of the blackbody component $\sim\pi$ out of phase with that of the power law are required to reproduce the data. Figure \begin{figure}
\includegraphics[angle=90,width=0.5\textwidth]{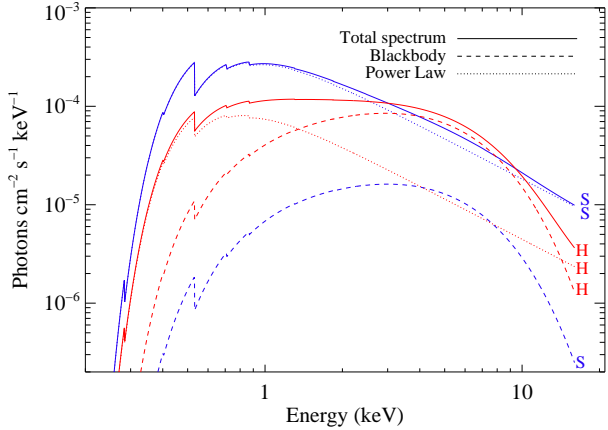}
\caption{Figure shows the ''hardest'' (red) and ''softest'' (blue) states of the spectrum, according to the simulations, and the model components.}\label{fig:minmax}
\end{figure} \ref{fig:minmax} shows the simulated spectrum and model components of Swift~J045106.8-694803 at the hardness ratio maximum (hardest) and minimum (softest). The pulsed fraction of the power law and blackbody components are consistent at $0.6\pm0.2$ and $0.7\pm0.3$ respectively.

Since the blackbody component varies with rotation and can be inferred to have a small (R\textless R$_{NS}$) emitting size from $L_X$ and $kT_{BB}$, it is possible that this region is a hot, magnetic polar cap on the neutron star surface. By adopting a model for the hot spot emission and fitting this model to the pulse profile, the geometry of the system can be constrained, in particular, the angle between the rotation and magnetic axes $\alpha$ and the angle between the rotation axis and line-of-sight $\zeta$.

\begin{figure}
\includegraphics[width=0.5\textwidth]{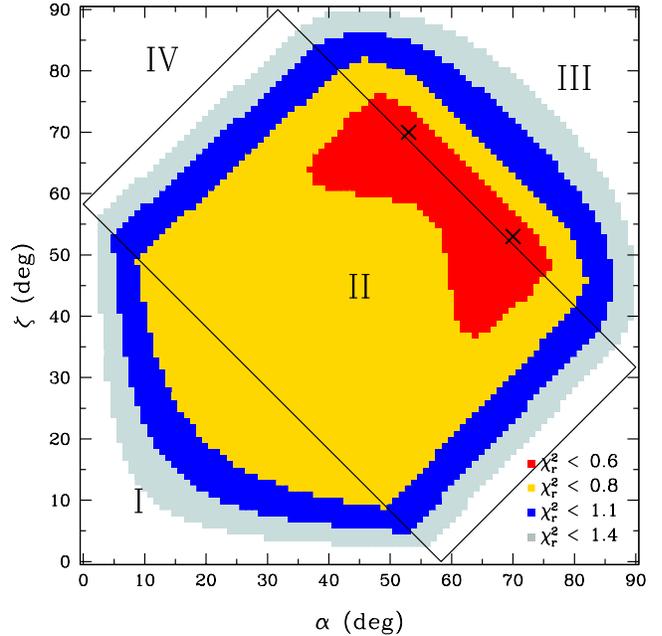}
\caption{$\chi^2_r$ contours of the fit to the blackbody pulse profile for the angle between rotation and magnetic axes, $\alpha$, and angle between rotation axis and
line-of-sight, $\zeta$. The 90\% confidence contour corresponds to a $\chi_r^2=1.1$. Crosses indicate the best-fit values. The four classes (\textsc{i}--\textsc{iv}) are defined in the text.}\label{fig:contours}
\end{figure}

Proper modelling requires detailed knowledge of the magnetic field and temperature distributions over the neutron star surface and is beyond the scope of this work (see, e.g., \citealt{Ho07}). Nevertheless useful insights can still be easily obtained using a simple model: two antipodal hot spots that emit as a blackbody and have a beam pattern (i.e., angular dependence) which is isotropic and has no energy-dependence. For each $(\alpha,\zeta)$, the pulse profile is calculated using the analytic approximation of \citep{Beloborodov02} to the exact relation given in \citet{Pechenick83} which accounts for gravitational light-bending. The pulse profiles are degenerate in the two angles, i.e., $(\alpha,\zeta)$ and $(\zeta,\alpha)$ produce the same profile. A neutron star mass $M_{\mathrm{NS}}=1.4\,M_\odot$ and radius $R_{\mathrm{NS}}=12$~km are assumed. These model pulse profiles are then fit to the blackbody pulse profile (see panel (3) of Fig. \ref{fig:sims}), allowing the phase and amplitude to vary. Shaded regions for $\chi_r^2$ (for 8 degrees of freedom) values are shown in Fig. \ref{fig:contours}.

For this emission model, the $\alpha-\zeta$ parameter space can be divided into four regions which correspond to four classes defined in \citep{Beloborodov02}: Class \textsc{i} is where one pole is visible all the time, the second pole never, class \textsc{ii} is where one pole is visible all the time and the second pole some of the time, class \textsc{iii} is where both spots are seen some of the time and class \textsc{iv} is where both spots are always seen. A geometry in class IV is immediately ruled out as it requires the blackbody flux to be constant. Similarly, the out of phase pulsations of the power law (interpreted as the accretion column) and the blackbody suggests we can also rule rule out a geometry in class I, if the accretion column is located just above the neutron star surface since it will always be eclipsed by the neutron star. The results from the fitting suggest that we are seeing both of the neutron star poles during a rotation of the neutron star with best-fit values for the angles of ($\alpha$,$\zeta$)=(53,70) with a $\chi^2$ of 4.32 for 8 dof.

\section{Discussion}

A soft excess is a common feature in the X-ray spectra of BeXRBs. It is hypothesised that a soft excess is in fact present in all BeXRB spectra, though not always detected due to the high intrinsic absorption and flux of some sources. It is thought to originate from the inner radius of an accretion disc surrounding the neutron star \citep{Hickox04}, however the majority of the blackbody temperatures reported are a factor $\sim$10 lower than that found here (e.g. \citealt{Sturm12}). The temperature and flux of the blackbody component detected in this observation of Swift~J045106.8-694803 ($kT_{BB}=1.8_{-0.3}^{+0.2}$~keV, $f_{X,BB}=1.3\pm0.8\times10^{-12}$~erg~cm$^{-2}$~s$^{-1}$) along with a distance to the LMC of $50.6\pm2.1$~kpc \citep{Bananos11} implies a blackbody radius of $R_{BB}$=0.5$\pm$0.2~km, calculated using the formula $R_{BB}=\sqrt{L_X/(4\pi\sigma T^4)}$. All errors represent the 90\% confidence limits. This is less than the radius of a neutron star and so emission from the entire accretion disc is clearly not the origin of this excess.

In the last 10 years a smaller subset of HMXBs have been discovered which are reported to have a ``hot'' ($kT_{BB}>1$~keV) thermal excess (see Table \ref{tab:BeX}). These all have $R_{BB}\lesssim$1~km, suggesting emission from the neutron star polar cap. Interestingly these systems are characterised by long pulse periods (P\textgreater100~s) and persistent X-ray emission, much like Swift~J045106.8-694803. Could this be a selection effect? The ability to detect pulsations in a given observation decreases with increasing pulse period. This could lead to observers requesting longer observations of the longer pulse period pulsars and thus having an greater number of counts in the source spectrum. This in turn would allow fainter spectral components to be detected. However, there are several pulsars with both short and long periods that have been observed with a similar or greater number of total counts than seen here which have not shown any evidence for this spectral component (see \citealt{Haberl08}, \citealt{Sturm12} and \citealt{Townsend11}, for recent examples with \emph{XMM}). As such we conclude that this cannot solely be an observational bias.

Thermal emission is often observed to have the greatest contribution to the soft energy band. This does not appear to be the case in the spectrum of Swift~J045106.8-694803, as the blackbody component contributes $\sim50$\% of the total flux of the source at energies $\gtrsim4$~keV. Lomb-Scargle analysis of the 4.5-10.0 keV light curve shows no evidence for pulsations, which seems to contradict the hypothesis that the thermal emission originates from the polar cap. However, simulations of the spectrum's model components suggest that the lack of pulsations is the result of the two components pulsating out phase. Figure \ref{fig:minmax} shows the ``hard'' and ``soft'' spectra of Swift~J045106.8-694803. Despite similar levels of variation in both components, the overall spectrum varies very little above $\sim3$~keV. This reflects the results of the Lomb-Scargle analysis of the higher energy light curves and also explains the reduction in the pulsed fraction at higher energies, which is usually observed to increase with both increasing energy and decreasing source flux as the regions emitting the X-rays become more compact \citep{Lutovinov08}. Above $\sim$10~keV, the non-thermal component once again dominates the X-ray spectrum. If this hypothesis is correct, then pulsations should once more be detectable at higher energies.

The decomposition of the spectral components has allowed us to demonstrate how the geometry of the neutron star could be constrained should better data become available. The current values of $\alpha$ and $\zeta$ are approximate since we assumed a simple blackbody emission model. More sophisticated modelling is not warranted at this time given the relatively large uncertainties of the pulse profile. Detailed modelling of deeper observations, with better signal-to-noise, could provide an independent measurement of the magnetic field; furthermore, future polarization studies could even break the degeneracy between $\alpha$ and $\zeta$.

\citet{Klus12} use the pulse period determined in this work along with data from \emph{Swift} and \emph{RXTE} to show that Swift~J045106.8-694803 has a magnetic field above the quantum critical value. Two other accreting pulsars are known with $\dot{P}$ values consistent with a super stong magnetic field, 4U2206+54 \citep{Reig12} and SXP~1062 \citep{Turolla12,Henault11}. Intriguingly, both of these sources are members of the hot thermal excess population (see Table \ref{tab:BeX}) which suggests a possible link between these two phenomena (although SXP~1062 is also surrounded by a supernova remnant with $kT_{BB}=0.23\pm0.05$~keV possibly adding to the thermal excess, \citealt{Haberl12a}). 

The link between the magnetic field and spin period of a neutron star is well known (e.g. \citealt{Shapiro83}). For isolated pulsars, the relationship is $B\propto (\dot{P}P)^{1/2}$ and is due to emission of magnetic dipole radiation. For accretion powered pulsars, the torque experienced due to accretion is much stronger than the dipole spin down torque and the relationship is $B\propto P^{7/6}$ if the neutron star is spinning at its equilibrium period (i.e. $\dot{P}=0$). If the neutron stars are not in spin equilibrium, the relationship is more complex (e.g. \citealt{Ghosh79} $B\propto (-\dot{P}/P^2)^{7/2}$). The ``standard'' relation for radio pulsars also links the neutron star polar cap size to its spin period, $\Theta\propto P^{-1/2}$, i.e. longer period pulsars have small caps. If this is extended to the accretion powered pulsars, the relationships above suggest that higher magnetic fields would be found in pulsars with longer periods and imply smaller polar caps.

\section{Conclusion}

We have presented detailed analysis of a recent \emph{XMM-Newton} ToO observation of the BeXRB Swift~J045106.8-694803. The period was determined to be 168.5$\pm$0.2~s as of MJD = 56125.0. The pulsed fraction decreases with increasing energy, with no detection of the period at energies \textgreater4.5~keV. The X-ray spectrum is adequately represented by two models, a single component continuum model (an absorbed powerlaw) and a two component continuum model (an absorbed powerlaw and blackbody). The extra blackbody component, though just formally significant, is not necessary for an acceptable fit to the spectrum and the parameters of the \emph{phabs*vphabs*powerlaw} model are consistent with those reported for other BeXRBs. However, it is difficult to explain the dramatic decrease in the pulsations with increasing energy with a single component model. A two component model, with anticorrelated pulsations, can account for this behaviour and the anticorrelation of the hardness ratio and pulse profile and implies $\alpha$ and $\zeta$ values of $\sim53\degree$ and $\sim70\degree$.

The high temperature of the blackbody ($kT_{BB}=1.8_{-0.3}^{+0.2}$~keV) implies a radius of blackbody radius of $R_{BB}=0.5\pm0.2$~km, and is attributed to the polar cap of the neutron star. This is not the first detection of a hot thermal excess in the X-ray spectra of HMXBs (see Table \ref{tab:BeX} for recent examples) and interestingly Swift~J045106.8-694803 shares common characteristics with these sources including persistent low level X-ray emission and a long pulse period (P\textgreater100~s). If confirmed to be the latest member of this emerging class, it would be the brightest and shortest period source. 

Interestingly, two of the other sources listed in Table \ref{tab:BeX} also have high $\dot{P}$ values, indicating a strong magnetic field. Whilst based on a small sample, this could suggest that there is a link between a hot thermal excess and the magnetic field strength. Further monitoring of the pulse periods of these sources as well as the temperatures of their thermal components could reveal if this is causal or coincidence. Should the blackbody in Swift~J045106.8-694803 exist, we predict that above 10~keV the pulse period should once more be detectable as the power law becomes the dominant component in the X-ray spectrum once again.

Most of the known X-ray pulsars in the Small and Large Magellanic cloud have been detected with \emph{RXTE} during a $\sim$10~yr monitoring campaign \citep{Galache08}. \emph{RXTE} has a limited response below 2~keV and this particular pulsar, with its soft pulses and low level emission, would not have been detected unless it went into outburst. Detailed analysis of the \emph{XMM} survey of the SMC \citep{Haberl12b} could reveal a further population of these softly pulsating HMXBs.

\section*{Acknowledgments}

We would like to thank Nicola La Palombara for making the spectrum and best fit model of RX~J0440.9+4431 available to us. This paper is based on observations obtained with XMM-Newton, an ESA science mission with instruments and contributions directly funded by ESA Member States and NASA. This research has made use of the NASA/IPAC Extragalactic Database (NED) which is operated by the Jet Propulsion Laboratory, California Institute of Technology, under contract with the National Aeronautics and Space Administration. ESB acknowledges support from a Science and Technology Facilities Council (STFC) studentship. WCGH appreciates the use of the computer facilities at the Kavli Institute for Particle Astrophysics and Cosmology and acknowledges support from STFC in the UK.

\bsp

\label{lastpage}

\end{document}